**Running Title: Joo *et al.*: Sensitivity Analysis of NF-κB signaling network**

# Title: Sensitivity analysis of a computational model of the IKK-NF-κB-IκBα-A20 signal transduction network

Jaewook Joo [1], Steve Plimpton [2], Shawn Martin [2], Laura Swiler [3], Jean-Loup Faulon [1]

[1] Computational Biosciences Dept., Sandia National Laboratories, PO Box 5800 Albuquerque, NM 87185-1413
[2] Computational Biology Dept., Sandia National Laboratories, PO Box 5800 Albuquerque, NM 87185-1412
[3] Optimization and Uncertainty Estimation Dept., Sandia National Laboratories, PO Box 5800 Albuquerque, NM 87185-1318

Mailing address: Computational Biosciences Dept., Sandia National Laboratories, PO Box 5800 Albuquerque, NM 87185-1413; Phone: 505-284-6766;
Email: jjoo@sandia.gov

## Abstract

The NF-κB signaling network plays an important role in many different compartments of the immune system during immune activation. Using a computational model of the NF-κB signaling network involving two negative regulators, IκBα and A20, we performed sensitivity analyses with three different sampling methods and present a ranking of the kinetic rate variables by the strength of their influence on the NF-κB signaling response. We also present a classification of temporal response profiles of nuclear NF-κB concentration into six clusters, which can be regrouped to three biologically relevant clusters. Lastly, based upon the ranking, we constructed a reduced network of the IKK-NF-κB-IκBα-A20 signal transduction.

## I. Introduction

NF-κB is a stimulus-responsive pleiotropic transcription activator and plays a significant role in various parts of the immune system during differentiation of immune cells, development of lymphoid organs, and immune activation [1]. Upon stimulation by LPS, TNFα, or UV irradiation, the NF-κB transcriptional activator is shuttled into the nucleus, initiating transcription of target genes responsible for inflammatory cytokines, anti-apoptic molecules, and NF-κB signal termination. NF-κB shuttling between nucleus and cytoplasm is regulated by the IKK-NF-κB-IκBα-A20 signaling module, which consists mainly of four proteins: IκBα, IKK (IκB kinase), A20, and NF-κB [1, 2, 3, 4, 5].

Recent computational models of NF-κB signal transduction have greatly enhanced our understanding of the underlying (negative regulation) mechanisms of NF-κB signaling and are being corroborated by experimental measurements. Hoffmann *et al.* [2] demonstrated that IκBα is responsible for strong negative feedback in NF-κB response which results in oscillatory shuttling of NF-κB transcription activator between cytoplasm and nucleus, whereas IκBβ and IκBε reduce the oscillation magnitudes. Nelson *et al* [6] reported a groundbreaking observation of sustained oscillations of NF-κB shuttling within a single cell, which drew attention to the possible cellular mechanism and functionality of the oscillatory pattern of NF-κB signaling [6, 7, 8, 9]. In addition to IκB isoforms with their negative regulatory role, A20, a cytoplasmic ubiquitin-modifying enzyme [5] is also required for termination of NF-κB activity, thus limiting TNFα-induced [4] or LPS-induced inflammation [5]. Lipniacki *et al.* [10, 11] modified the NF-κB model of Hoffmann *et al*. [2, 3] by adding A20 as an additional NF-κB activity terminator, and excluding IκBβ and IκBε, thereby successfully reproducing experimentally observed NF-κB signaling response for A20-deficient cells [4].

Sensitivity analysis allows the identification of the most significant kinetic reactions which control the dynamic patterns of NF-κB response, i.e., oscillation of NF-κB shuttling. Ihekwaba *et al*. [12, 13] performed a sensitivity analysis on a simplified version of Hoffmann's NF-κB signaling model that considers only IκBα, NF-κB, IKK, and their complexes. Due to the computational cost of sampling 64 kinetic rate variables, their sensitivity analysis was limited to a single-variable variation [12] and at most a pair-wise modulation of 9 kinetic rate variables presumed to be of high importance [13].

For this paper, we use Lipniacki *et al.*'s model of the IKK-NFκB-IκBα-A20 signal transduction network [10, 11] in Fig. 1. This model includes two negative regulators, IκBα and A20. We performed sensitivity analysis on the network using three different sampling methods: single-variable variation, Orthogonal Array sampling, and Latin Hypercube sampling. The sensitivity analyses of the resulting temporal profiles of nuclear NF-κB response enable an importance ranking of all the kinetic rate variables in the model. We also present a classification of the resulting temporal profiles of nuclear NF-κB concentration into six clusters, which are regrouped further into three biologically relevant clusters. We propose a reduced network of IKK-NFκB-IκBα-A20 signal transduction based on these critical kinetic rate variables.

## II. Network Model and Dynamics

### 1. IKK-NF-κB-IκBα-A20 network model:

The NF-κB signaling network model proposed in [10, 11] and presented in Fig. 1 involves the kinetics of IKK, NF-κB, A20, IκBα, their complexes, mRNA transcripts of A20 and IκBα, and translocation/shuttling of NF-κB and IκBα between nucleus and cytoplasm. The regulatory module has two activators IKK and NF-κB, and two inhibitors A20 and IκBα. In resting cells, unphosphorylated IκBα binds to NF-κB and sequesters NF-κB in an inactive form, namely IκBα-NF-κB, in the cytoplasm. In the presence of an extracellular stimulus such as by TNF or LPS, IKK is transformed into its active/phosphorylated form and is capable of phosphorylating IκBα, leading to ubiquitin-assisted proteolysis of IκBα. As a result of degradation of IκBα, free NF-κB enters the nucleus and upregulates transcription of the two inhibitors IκBα and A20. The newly synthesized IκBα inhibits NF-κB activity by sequestering it in the cytoplasm while A20 negatively regulates IKK activity by transforming IKK into an inactive form, in which IKK is no longer capable of phosphorylating IκBα. As NF-κB, IκBα and their complexes are translocated from cytoplasm to nucleus, their concentrations change by a factor equal to the volume ratio of cytoplasm to nucleus, namely Kv.

The IKK-NF-κB-IκBα-A20 network model in Fig. 1 and Table I is readily translated into a set of ordinary differential equations with 15 dependent variables, 25 kinetic rate variables, and the initial cytoplasmic concentration of NF-κB as an initial condition. The total concentration of NF-κB, summed across all involved complexes, remains constant. Runge-Kutta 4$^{th}$ order method is used to numerically solve the ordinary differential equations.

2. **Dynamic Features:**

We characterize a temporal profile of nuclear NF-κB with its four salient dynamic characteristics: (1) period, defined as the average time interval between all pairs of adjacent maxima and adjacent minima; (2) damping constant, defined as the average slope between all pairs of adjacent maxima and adjacent minima; (3) steady state amplitude, defined as the average nuclear NF-κB level between 9 and 10 hours post stimulation; and (4) phase, defined as the time-point of the first maximum of nuclear NF-κB. These features are shown graphically in Fig. 2.

Each of nuclear NF-κB profile is quantified with four numeric dynamic features as the kinetic rate variables vary. Each of four dynamic features is normalized with its own maximum score, i.e., each of the four normalized dynamic feature scores is always between zero and one, and the nuclear NF-κB response is quantitatively represented by the average (equally-weighted) score of these four normalized dynamic feature scores. For reference, we also calculated a distance measure (5), defined as a sum of absolute distance between the reference temporal profile of nuclear NF-κB in Fig. 2 and a perturbed temporal profile at discrete time points.

3. **Correlation Coefficient:**

The correlation coefficient between a pair of vectors *X* and *Y*, is defined as:

$$Corr(X,Y) = \frac{\sum_{i=1}^{N}(X_i - \overline{X})\cdot(Y_i - \overline{Y})}{\sqrt{\sum_{i=1}^{N}(X_i - \overline{X})^2 \cdot \sum_{i=1}^{N}(Y_i - \overline{Y})^2}},$$

where $\overline{X}$ and $\overline{Y}$ are the averages of *X* and *Y*, respectively. In this analysis, *X* is a vector of *N* values that a particular input kinetic rate variable was set to, and *N* simulations of the model were performed. The resulting *N* temporal profiles of nuclear NF-κB were scored according to one of the 4 dynamic feature metrics, and *Y* is the vector of *N* scores. The correlation coefficient is always a number between -1 and +1. If the correlation coefficient is close to +1 (or -1), then there is a strong positive (or negative) linear relationship between X and Y. We utilize the magnitude of the correlation coefficient to investigate whether the relationship between a kinetic rate variable and a dynamic feature of temporal profile of nuclear NF-κB is statistically significant.

## III. Methods

### 1. Sampling Techniques:

Various sampling techniques were employed to address the need for sensitivity analysis or uncertainty assessment for the computational model of the NF-κB signaling pathway. The techniques are particularly useful for the investigation of how NF-κB response varies when the kinetic rate variables of the model vary according to some assumed (joint probability) distribution.

#### a) Single-variable variation:

A single-variable variation scheme allows one to vary a single kinetic rate variable of the computational model while keeping the rest of kinetic rate variables at their fixed nominal values. We changed each kinetic rate variable to one tenth or four times of its nominal value (see Table I) and then measured the impact of the kinetic rate variable on the nuclear NF-κB signaling response.

This method is one of the simplest and computationally least demanding sampling procedures. Its disadvantage, however, outweighs its merit. This sampling scheme samples a kinetic rate variable along a line segment in a 25-dimensional space for 25 kinetic rate variables and leaves almost the entire space unexplored. We compensate for this weakness with two additional sampling methods.

#### b) Orthogonal Array (OA) sampling

The OA sampling technique enables one to calculate the main effect of simultaneous variations of all kinetic rate variables with a relatively small number of sample points. The OA matrix consists of $N$ orthogonal vectors, each of which represents a single test set conveying unique information about input kinetic rate variables, i.e., a vector of a unique combination of discrete values of the 25 input kinetic rate variables. In our case each input kinetic rate variable can be one of three discrete values: a "reference" level denotes the nominal value of an individual kinetic rate variable; a "high" level denotes twice the nominal value; and a "low" level denotes one fifth of the nominal value. The "high" and "low" levels were chosen such that each variable changes within a biologically feasible range of a factor of 10.

The OA matrix is restricted such that (i) any pair of columns should be orthogonal, (ii) they must contain all possible combinations of three levels an equal number of times, and (iii) the number of sample points must be a multiple of the square of the number of the levels, e.g., , 9. We used an OA matrix with 25 kinetic rate variables, 81 sample points, and 3 levels. For details see [17]. After sampling 25 kinetic rate variables according to the OA sampling method, we measure the mean response of NF-κB signaling with respect to one of three levels, i.e., the average dynamic feature score of nuclear NF-κB temporal profiles when the level of an individual kinetic rate variable is at "reference" level, or when it is at "high" level, or when it is at "low" level, respectively. If the mean response of the NF-κB signaling response for one level of a kinetic rate variable is statistically significantly different from that for one of the other two levels, that kinetic rate variable is said to have a significant effect on the NF-κB signaling response.

c) **Latin Hypercube sampling (LHS):**

LHS is a constrained Monte Carlo sampling scheme. The Monte Carlo sampling scheme is a conventional approach and a common choice for the uncertainty assessment of a computational model. By sampling repeatedly from the assumed joint probability function of the input variables and evaluating the response for each sample, the distribution of the response of the computer model can be estimated. This approach yields reasonable estimates for the distribution of the response if the number of samples is quite large. However, since a large sample size requires a large number of computations from the computer model (a potentially very large computational expense), an alternative approach, Latin Hypercube sampling, can be used.

LHS yields more precise estimates with a smaller number of samples, and is designed to address the above concern [14]. Suppose that the computer model has K kinetic rate variables and we want N samples. LHS selects N different values from each of K kinetic rate variables such that the range of each variable is divided into N non-overlapping intervals on the basis of equal probability. One value from each interval is selected at random with respect to the assumed probability density in the interval. The N values thus obtained for the first kinetic rate variable are paired in a random manner (equally likely combinations) with the N values of the second kinetic rate variable. These N pairs are combined in a random manner with the N values of the third kinetic rate variable to form N triplets, and so on, until N K-tuplets are formed. These N K-tuplets are the same as the N K-dimensional input vectors where the *ith* input vector contains specific values of each of the K kinetic rate variables to be used on the *ith* run of the computer model [14].

For the sensitivity analysis of the NF-κB signaling model we sampled N=1000 sets for K=25 kinetic rate variables. Since the distribution of each of K kinetic rate variables is unknown, we assume a simple yet practical distribution for each of K kinetic rate variables: a uniform distribution with an interval, (0.2Xi, 1.8Xi), where Xi is a nominal value of the *ith* kinetic rate variable that was used in [10, 11]. The pair-wise correlation between samples is very small.

Taking into account the number of samples and the reliability of each of three sampling methods, the LHS is the most reliable sampling scheme whereas the OA is the most efficient and sophisticated sampling method (OA sampling enables us to obtain mean responses of the computational model with as few samples as possible). The single-variable variation method is least reliable because of its very poor representation of the sample space.

2. **K-means Clustering**:

K-means is an unsupervised learning algorithm for partitioning a given data set into K clusters [15]. It defines K centroids (one for each cluster), and minimizes the sum of the squared distances of each data point to the nearest centroid. The minimization is performed using a stochastic iterative method.

The main difficulty in using K-means is the choice of the number (K) of clusters. This choice can be made using a priori knowledge or, more often, a posteriori error analysis. Such error analysis is often performed using 10-fold cross validation. Cross validation is a common procedure wherein a dataset is divided into ten equal subsets. For each (test) subset, K-means is performed on the nine remaining (training) subsets. The sum of the squared distances of each data point in the test subset is computed as an indicator of the accuracy of the clustering. This quantity is averaged over the ten test subsets to provide a measure of cluster quality (lower is better).

## III. Results

### 1. Sensitivity Analysis

In this subsection we sample the kinetic rate variables of the NF-κB signaling network according to the three sampling methods: single-variable variation, LHS, and OA sampling. We quantify the nuclear NF-κB temporal profiles and rank the kinetic rate variables in order of their influence on the NF-κB signaling response.

#### a. Sensitivity analysis with single-variable variation:

We performed the sensitivity analysis with single-variable variation and ranked 25 kinetic rate variables in order of their influence on the nuclear NF-κB response. Each of 25 kinetic rate variables is perturbed according to the single-variable variation scheme. The numerically simulated nuclear NF-κB profiles are quantified with four dynamic feature scores. For each dynamic feature, its deviation from the same dynamical feature score in the reference NF-κB profile in Fig, 2 is assessed and normalized with its maximum deviation score, yielding a score between zero and one. We took an average of two normalized deviation scores for each of the four dynamical features and for each of the 25 kinetic rate variables. In addition, assigning an equal weight (0.25) to each of four normalized deviation scores, we calculated the equally-weighted average deviation score from four deviation scores for each of 25 kinetic rate variables.

As shown in Fig.3, the kinetic rate variables are ranked primarily based on the equally-weighted deviation scores. The equally-weighted deviation score agree well with the absolute distance as well as the individual dynamic feature scores for each of kinetic rate variables. We also changed the variation size of a single kinetic rate variable from "one tenth and four times" to "one fifth and twice" of its nominal value and confirmed that the ranks of the kinetic rate variables were unchanged.

A cutoff threshold with an equally-weighted score of 0.2 partitions the 25 kinetic rate variables into two groups: one with significant impact and the other with negligible impact to nuclear NF-κB response. This cutoff choice was motivated by the abrupt change of equally-weighted scores in Fig. 3. The categorization is also guided by the list of "statistically significant" kinetic rate variables from the OA-sensitivity analysis which is discussed below. Even though sensitivity analysis with single-variable variation is not as reliable as the other two methods we discuss, the two resulting groups are consistent with the lists resulting from LHS and OA sampling as presented in Table II.

The kinetic rate variables that are classified as most important by this sensitivity analysis are mostly related to transcription and translation of IκBα and A20. The transcriptional activity is known to be slow and hence greatly affected by noise-induced variation of the transcription-related kinetic rate variables.

We also applied the single variable variation scheme to the computational model of NF-κB proposed by Hoffmann *et al* [3] and found a list of important/relevant kinetic rate variables consistent with the list provided by other sensitivity analysis with single variable variation in [12]. The dissociations of protein complexes, which are included in the computational model of Hoffmann *et al.* [3], e.g., IKKa-IκBα-NF-κB → IKKa-IκBα + NF-κB, IKKa-IκBα-NF-κB → IKKa + IκBα-NF-κB, and IκBα-NF-κB → IκBα + NF-κB, make negligible contribution to the NF-κB response and this justifies the model of Lipniacki *et al.* [10, 11], which discards all dissociations of protein complexes.

b. **Sensitivity analysis with Orthogonal Array sampling:**

After sampling 25 kinetic rate variables by the OA sampling method, we measured the mean response of NF-κB signaling. This value was computed using the equally-weighted and normalized dynamic feature scores for each of the three levels of individual kinetic rate variables as shown in Fig. 4.

A kinetic rate variable is categorized as having a significant impact on the NF-κB signaling response, when the mean response of the NF-κB signaling for an individual kinetic rate variable at one level is significantly different (in the statistical sense) from the mean response at one of the other two levels. As shown in Fig. 4, the mean response of the NF-κB signaling at the "high" level of any given kinetic rate variable is not statistically significantly different from the mean response at the "reference" level of that kinetic rate variable (the only exception was the C3a kinetic rate variable). This indicates that the nuclear NF-κB response is more significantly influenced when a kinetic rate variable value decreases from its nominal value than when it increases.

The kinetic rate variables whose variation generates significantly different mean responses of NF-κB signaling are listed on the left side of Fig. 4 and can be categorized into 6 biologically distinct groups as follows: (i) volume ratio of cytoplasm to nucleus that affects nuclear NF-κB concentration (Kv); (ii) synthesis and degradation of mRNA IκBα and mRNA A20 that negatively regulates NF-κB signaling (C1a, C3a, C3); (iii) activation and inactivation of IKK that initiate and continuously maintain NF-κB signaling (Kprod, K2); (iv) association of IKKa and IκBα-NF-κB that leads to proteolysis of IκBα and liberation of NF-κB for transcription activity (A3); (v) spontaneous degradation of protein IκBα from IκBα-NF-κB that leads to liberation of NF-κB (C6a and T1); (vi) translocation of NF-κB into the nucleus (I1).

c. **Sensitivity analysis with Latin Hypercube sampling:**

We used LHS to obtain a thousand samples of 25 kinetic rate variables taken uniformly from a 25-dimensional hypercube. The hypercube was defined as a product of 25 intervals, (0.2Xi, 1.8Xi), where Xi is the *ith* nominal kinetic rate variable found in the literature [2, 3, 10, 11]. We generated the nuclear NF-κB temporal profiles by solving a system of ODE equations and computed the correlation coefficients between individual kinetic rate variables and each of the four dynamic feature scores. For each of the 25 kinetic rate variables, we assigned an equal weight to the four correlation coefficients and used the Root Mean Square score (RMS) of the four correlation coefficients to place the 25 kinetic rate variables in order of RMS score from high to low. This ranking scheme corresponds to the order of influence on the NF-κB response and presented in Fig. 5.

Because the RMS score varies smoothly without an abrupt transition/change in Fig. 5, it is hard to introduce a cut-off RMS score and partition the 25 kinetic rate variables into two groups as before, namely important and unimportant kinetic rate variables. However, the partitioning resulting from a cutoff RMS score of 0.1 is consistent with the categorization produced by the OA sampling analysis, as shown in Table II.

The list of influential kinetic rate variables from the LHS-sensitivity analysis includes mRNA synthesis of IκBα (C1) as well as synthesis and degradation of protein IκBα and A20 (C4a, C4 ,C5) in addition to the kinetic rate variables classified as important from the OA-sensitivity analysis. Hence the complete list of influential kinetic rate variables are volume ratio of cytoplasm to nucleus (Kv), synthesis and degradation of mRNA and protein IκBα and A20 (C1, C3, C4, C5, C1a, C3a, C4a), activation and inactivation of IKK (Kprod, K2), association of IKKa and IκBα_NF-κB (A3), and translocation of NF-κB into nucleus (I1).

Both the RMS score and the absolute values of the correlation coefficients between the four dynamic features and individual kinetic rate variables are compared side-by-side in Fig. 5. Even though the RMS score and the scores from individual dynamic features have the similar trends (ups and downs) for some rate variables, they are not similar at all for other rate variables. As an example, the kinetic rate variable, I1, is very highly correlated with "phase" but is moderately correlated with other three dynamic features. If we assign more weight on "phase", then the ranking list will be different from one in Table II. Biological functionality of the four dynamic features is not known currently and hence the RMS score that weighs each of dynamic features equally is acceptable. However, as we gain more detailed knowledge about which of the four dynamic features are more biologically relevant, we can assign more weight to the more biologically relevant dynamic feature and consequently will obtain a new list of the influential kinetic rate variables.

2. **Clustering of the NF-κB signaling response:**

In this subsection, we analyze the nuclear NF-κB temporal profiles that the NF-κB signaling network generates in response to variation of the kinetic rate variables. We present a clustering analysis of 200 temporal profiles of nuclear NF-κB concentration, which were generated with the LHS-sampled kinetic rate variables according to a joint uniform distribution with 80% interval size for the 25 nominal kinetic rates in Table 1.

Many of the temporal profiles of the nuclear NF-κB concentration exhibit quasi-periodic behavior. Taking into account this quasi-periodicity, we preprocessed the time-series data with the Fast Fourier Transformation and applied K-means [15]. We used 10-fold cross validation to determine the appropriate number of clusters for the 200 time-series data points. According to our analysis, as shown in Fig. 6, a value of K=6 occurs at the "knee" of the curve in Fig. 6 and gives a small number of relatively accurate clusters.

For K=6, we obtained the six clusters shown in Fig. 7. These six clusters were identified using an unsupervised learning algorithm based on the nuclear NF-κB profiles, but they don't necessarily have distinctive/distinguishable physical or biological characteristics. To interpret the clusters, we grouped the six clusters into three distinguishable meta-clusters according to our understanding of the dynamical patterns of nuclear NF-κB concentration and assign a biological meaning to each of them: (1) the first meta-cluster consists of the curves in Fig. 7(c) and (e) and contains most of the single-peaked curves with a prominent first peak followed by a plateau; (2) the second meta-cluster consists of the curves in Fig. 7(b) and (f) and contains the curves with spiked oscillation, i.e., a 3-5 hour time lapse between the first peak followed by a swift fall to a zero level and subsequent oscillation; lastly (3) the third meta-cluster consists of the curves in Fig. 7(a) and (d) which contains oscillatory curves with the first peak immediately followed by the second peak.

The biological meaning of the three meta-clusters may be interpreted as follows. For the first meta-cluster of temporal profiles of nuclear NF-κB, the liberated NF-κB proteins upon stimulation migrate into the nucleus and their nuclear concentration doesn't decrease substantially for a prolonged time, resulting in unregulated hyper-inflammation. For the second meta-cluster the liberated NF-κB proteins rush into the nucleus, generate the first wave of immune response, are depleted from the nucleus for next 3-5 hours, and induce the second wave of weaker immune response, presumably only when the external stimulus still exists. The temporal behavior of the second cluster represents the most tightly regulated NF-κB signaling response, which monitors the change in external stimulus and minimizes the unnecessary inflammation-induced damage to the host. The third meta-cluster has the combined/mixed effect of the previous two clusters.

### 3. Reduction of the IKK-NF-κB-IκBα-A20 signal transduction network into a reduced network

In this subsection we compare the lists of the influential kinetic rate variables from the sensitivity analyses associated with the three sampling methods. Based on the results from the most reliable schemes, we construct a reduced network of IKK-NF-κB-IκBα-A20 signal transduction.

Table II presents a side-by side comparison of the three lists of the 25 kinetic rate variables using the three different ordering methods. (Low ranked variables are not included). Apart from a couple of exceptions, three sensitivity analyses using unrelated sampling and assessment methods furnish mutually consistent lists. For example, eight out of 10 important kinetic rate variables from OA-sensitivity analysis are ranked highly in the ranking list from the LHS-sensitivity analysis. The list of top 12 kinetic rate variables from the sensitivity analysis with single-variable variation is in agreement with the list from the sensitivity analysis using LHS.

The OA-sensitivity analysis results were used to determine the partitioning criteria for the single-variable variation and LHS analyses. High consistency between the lists of influential kinetic rate variables produced by all three statistical sampling methods in Table II was achieved with a RMS cutoff score of 0.1 for LHS and a deviation cutoff score of 0.2 for single-variable variation.

Using the binary classification of the kinetic rate variables from the sensitivity analyses with the LHS and the OA sampling, we further classified the important kinetic rate variables into two classes. When both the LHS and OA methods classify a kinetic rate variable as important, the kinetic rate variable is considered as "primarily important", whereas a kinetic rate variable is called "secondarily important" when only one of two schemes classify it as important. Primarily important kinetic rate variables are Kv, I1, C1a, C3a, C3, A3, K2, and Kprod whereas secondarily important kinetic rate variables are C1, C4, C5, C4a, C6a, and T1. Note that C6a and T1 are classified as important by the sensitivity analysis with the OA sampling but not included in the truncated list generated by the LHS-sensitivity analysis LHS and those two reactions are excluded from the reduced network in Fig. 8.

Finally we present the essential components (nodes) and interactions (edges) of the IKK-NF-κB-IκBα-A20 signaling module that critically govern the dynamics of the NF-κB response. In Fig 8 (a), insignificant kinetic reactions are marked with the dashed lines and the important kinetic reactions are marked with the solid lines. After removing insignificant reactions and components from the NF-κB signaling network, a few of the essential components are disconnected from the rest of the reduced network as shown in Fig. 8 (a). We make the essential components of the reduced network properly connected by allowing a few necessary yet insignificant interactions (such as K01, A1, T2) to be included in the reduced network. Assuming that two proteins, IKKn and IKKa-IκBα-NF-κB, are in a quasi-steady state due to their fast reactions, we removed IKKn and IKKa-IκBα-NF-κB from the network in Fig. 8 (a) and then renormalized K1 by K1Kprod/Kdeg. In Fig. 8(b) is presented the reduced network of the IKK-NF-κB-IκBα-A20 signaling module which consists of five proteins (IKK, NF-κB, NF-κBn, IκBα, A20), a heterodimer (IκBα-NF-κB), and two mRNAs (A20t and IκBαt). Because the total concentration of NF-κB, summed across all NF-κB involved complexes, is assumed to remain constant, one of three proteins, NF-κB, NF-κBn, and IκBα-NF-κB, is unnecessary and can be omitted from the reduced network. This reduced network includes all the relevant 15 interactions: NF-κB translocation (Kv, I1, K01); syntheses and degradation of mRNA and protein (C1, C3, C4, C5, C1a, C3a, C4a, C5a); formation of protein complex IκBα-NF-κB (A1); IKK-mediated IκBα degradation (A3); activation and inactivation of IKK (K1Kprod/Kdeg, K2). We conjecture that this reduced signaling network with these essential key components and interactions in Fig. 8 (b) can reproduce the essential dynamics of the full IKK-NF-κB-IκBα-A20 signaling network in Fig. 1.

## IV. Conclusion

We sampled 25 kinetic rate variables of the NF-κB signaling network according to three sampling methods, single-variable variation, Latin Hybercube sampling, and Orthogonal Array sampling. We quantified the nuclear NF-κB temporal profiles with respect to four dynamic features, and ranked the kinetic rate variables in order of their influence on the NF-κB signaling response. We also presented a classification of 200 temporal profiles of nuclear NF-κB concentration into six clusters, and then regrouped them further into three biologically relevant clusters. We made comparisons between the sensitivity analysis results from the three sampling methods. Based on results from the sensitivity analyses using Latin Hypercube and Orthogonal Array samplings, we constructed a reduced network of IKK-NF-κB-IκBα-A20 signal transduction.

The reduced network consists of 5 essential proteins (IKK, NF-κB, NF-κBn, IκBα, A20) and two mRNAs (A20t and IκBαt), along with 15 significant and necessary reactions. The reduced network model is similar to one of conceptual “cartoon-like” NF-κB signaling networks, which have been used only to illustrate and elucidate the key components and their primary reactions, but whose scientific importance has often been neglected because of the extreme simplicity with which they represent NF-κB signal transduction network. Our sensitivity analyses show that it is indeed plausible to represent the rich dynamics of detailed IKK-NF-κB-IκBα-A20 signal transduction networks such as those in [2, 3, 10, 11] with such a reduced network model. We also propose that the reduced network consisting of 7 components and 15 interactions can be further reduced to a minimal network model with only 4 components, namely IKK, NF-κB, IκBα, and A20. The further details about this reduction scheme of the full IKK-NF-κB- IκBα-A20 signaling network into the minimal network model and its computational and mathematical analysis will be published elsewhere [16].

## V. Acknowledgement

Sandia is a multi-program laboratory operated by Sandia Corporation for the United Sate Department of Energy under contract DE-AC94AL85000. The authors acknowledge the Laboratory Directed Research and Development program for funding.

Table I. List of kinetic rate variables, their units, and their nominal values for the computational model of the IKK-NF-κB-IκBα-A20 signaling network in [10, 11]

| Kinetic reactions | Kinetic rate variables | Unit | Nominal values |
|---|---|---|---|
| IκBα + NFκB → IκBα-NFκB | A1 | μM-1 s-1 | 0.5 |
| IκBαn + NFκBn → IκBαn-NFκBn | A1 | μM-1 s-1 | 0.5 |
| IKKa + IκBα → IKKa-IκBα | A2 | μM-1 s-1 | 0.2 |
| IKKa+IκBα-NFκB → IKKa-IκBα-NFκB | A3 | μM-1 s-1 | 1 |
| NFκBn → NFκBn + A20t | C1 | s-1 | 0.0000005 |
| 0 → A20t | C2 | μM s-1 | 0 |
| A20t → 0 | C3 | s-1 | 0.0004 |
| A20t → A20t + A20 | C4 | s-1 | 0.5 |
| A20 → 0 | C5 | s-1 | 0.0003 |
| NFκBn → NFκBn + IκBαt | C1a | s-1 | 0.0000005 |
| 0 → IκBαt | C2a | μM s-1 | 0 |
| IκBαt → 0 | C3a | s-1 | 0.0004 |
| IκBαt → IκBαt + IκBα | C4a | s-1 | 0.5 |
| IκBα, IκBαn → 0 | C5a | s-1 | 0.0001 |
| IκBα-NFκB → NFκB | C6a | s-1 | 0.00002 |
| IκBαn → IκBα | E1a | s-1 | 0.0005 |
| IκBαn-NFκBn → IκBα-NFκB | E2a | s-1 | 0.01 |
| NFκB → NFκBn | I1 | s-1 | 0.0025 |
| IκBα → IκBαn | I1a | s-1 | 0.001 |
| IKKn → IKKa | K1 | s-1 | 0.0025 |
| A20 +IKKa → A20 + IKKi | K2 | s-1 | 0.00000005 |
| IKKa → IKKi | K3 | s-1 | 0.0015 |
| IKKn, IKKa, IKKi → 0 | Kdeg | s-1 | 0.000125 |
| 0 → IKKn | Kprod | μM s-1 | 0.000025 |
| Volume ratio of nucleus to cytoplasm | Kv | 1 | 5 |
| IKKa-IκBα → IKKa | T1 | s-1 | 0.1 |
| IKKa-IκBα_NFκB → IKKa + NFκB | T2 | s-1 | 0.1 |
| Total NFκB concentration (NFκB+NFκBn) | | μM | 0.06 |

Table II. Ranking lists of kinetic rate variables in order of their influence on nuclear NF-κB response. These lists are obtained from sensitivity analyses using three different sampling schemes.

| Kinetic rate variables | Latin Hypercube sampling | Orthogonal Array sampling | Single-variable variation |
|---|---|---|---|
| Kv | 1 | Significant | 1 |
| I1 | 2 | Significant | 14 |
| C4a | 3 | Insignificant | 2 |
| C1a | 4 | Significant | 3 |
| C1 | 5 | Insignificant | 6 |
| A3 | 6 | Significant | 5 |
| K2 | 7 | Significant | 8 |
| Kprod | 8 | Significant | 13 |
| C3a | 9 | Significant | 4 |
| C4 | 10 | Insignificant | 7 |
| C3 | 11 | Significant | 9 |
| C5 | 12 | Insignificant | 10 |

Figure captions

Fig. 1. A network model of IKK-NF-κB-IκBα-A20 signal transduction network in [10, 11]. Squares represent proteins, hexagons represent mRNAs, black lines represent Protein-Protein interactions, blue lines represent protein synthesis, and red lines represent mRNA synthesis. Cytoplasm and nucleus is divided by a dashed line.

Fig. 2. Reference temporal profile of nuclear NF-κB and four dynamic features. This nuclear NF-κB temporal profile is simulated with the nominal kinetic rate values in Table I. Four dynamic features are period, damp, steady state, and phase.

Fig. 3. Sensitivity analysis using single-variable variation. Normalized score of deviation of each of the four dynamic patterns (period, damp, steady, phase) from the reference curve in Fig.2 are plotted against 25 kinetic rate variables.

Fig. 4. Sensitivity analysis using Orthogonal Array sampling. The mean responses (diamonds, circles, and squares) and the standard deviations (error bars) of the nuclear NF-KB are plotted against each of 25 kinetic rate variables. The mean response is defined as the mean score of equally weighted scores of four normalized dynamic feature scores of nuclear NF-κB response out of one third of N=81 samples for each of three levels of individual kinetic rate variables.

Fig. 5. Sensitivity analysis using Latin Hypercube sampling. For each of the 25 kinetic rate variables, absolute values and root mean square (RMS) score of four correlation coefficients are plotted. The 25 kinetic rate variables are placed in order of high RMS score from the left to the right.

Fig. 6. 10-fold cross validation using K-means clustering algorithm with different values of K. The K-means clustering is performed on 200 nuclear NF-κB temporal profiles are simulated with LHS –sampled kinetic rate variables.

Fig. 7. K-means clustering of 200 nuclear NF-κB temporal profiles into six clusters (a)-(f).

Fig. 8. Reduced network model for the IKK-NF-κB-IκBα-A20 signal transduction network. (a) Important reactions whose variation critically change the nuclear NF-κB response are marked with solid lines while insignificant interactions are denoted by dashed lines. (b) A reduced network of the IKK-NF-κB-IκBα-A20 signal transduction with 7 essential components and 15 reactions (IκBα-NF-κB can be omitted from the reduced network).

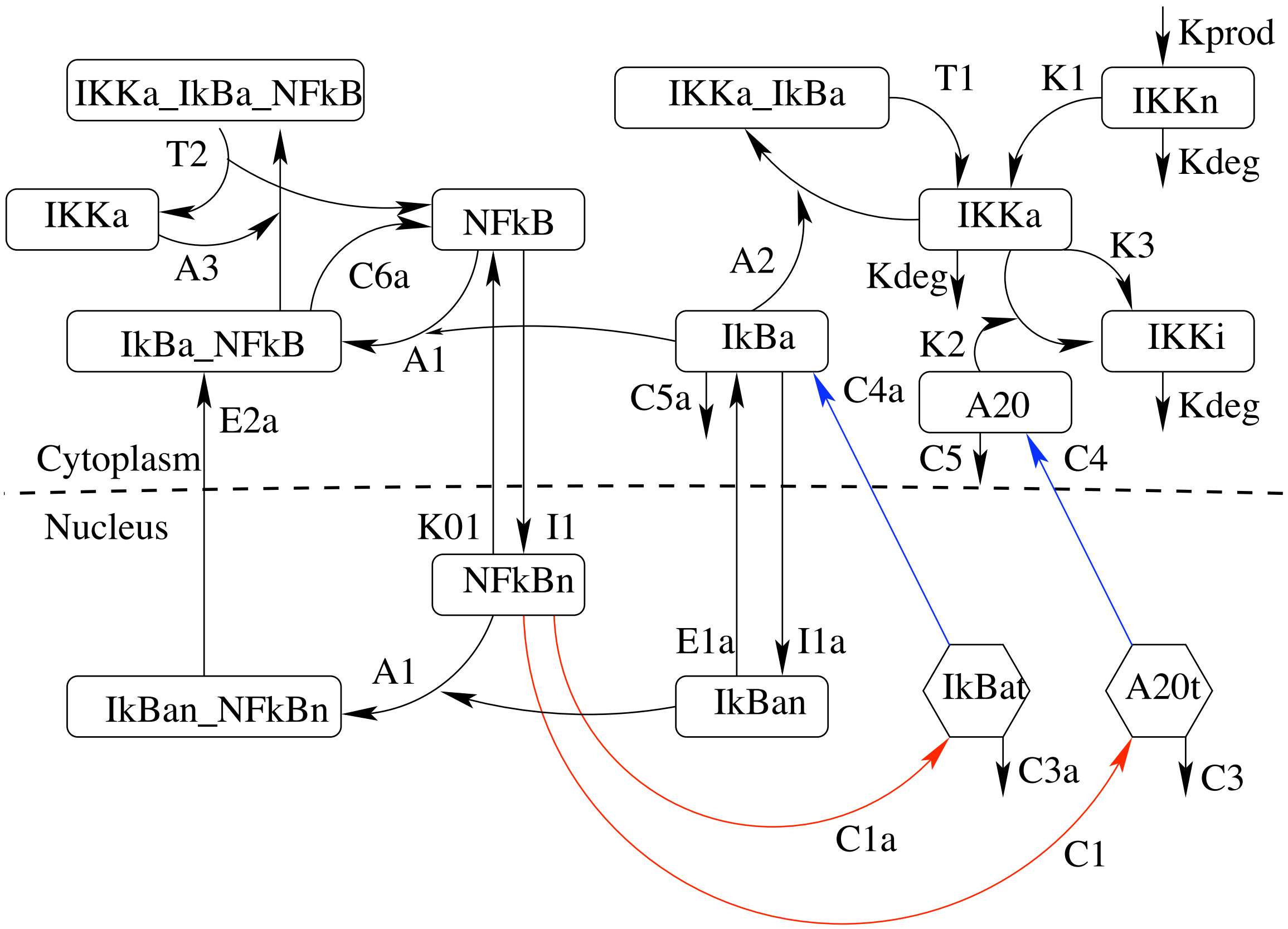

Figure 1

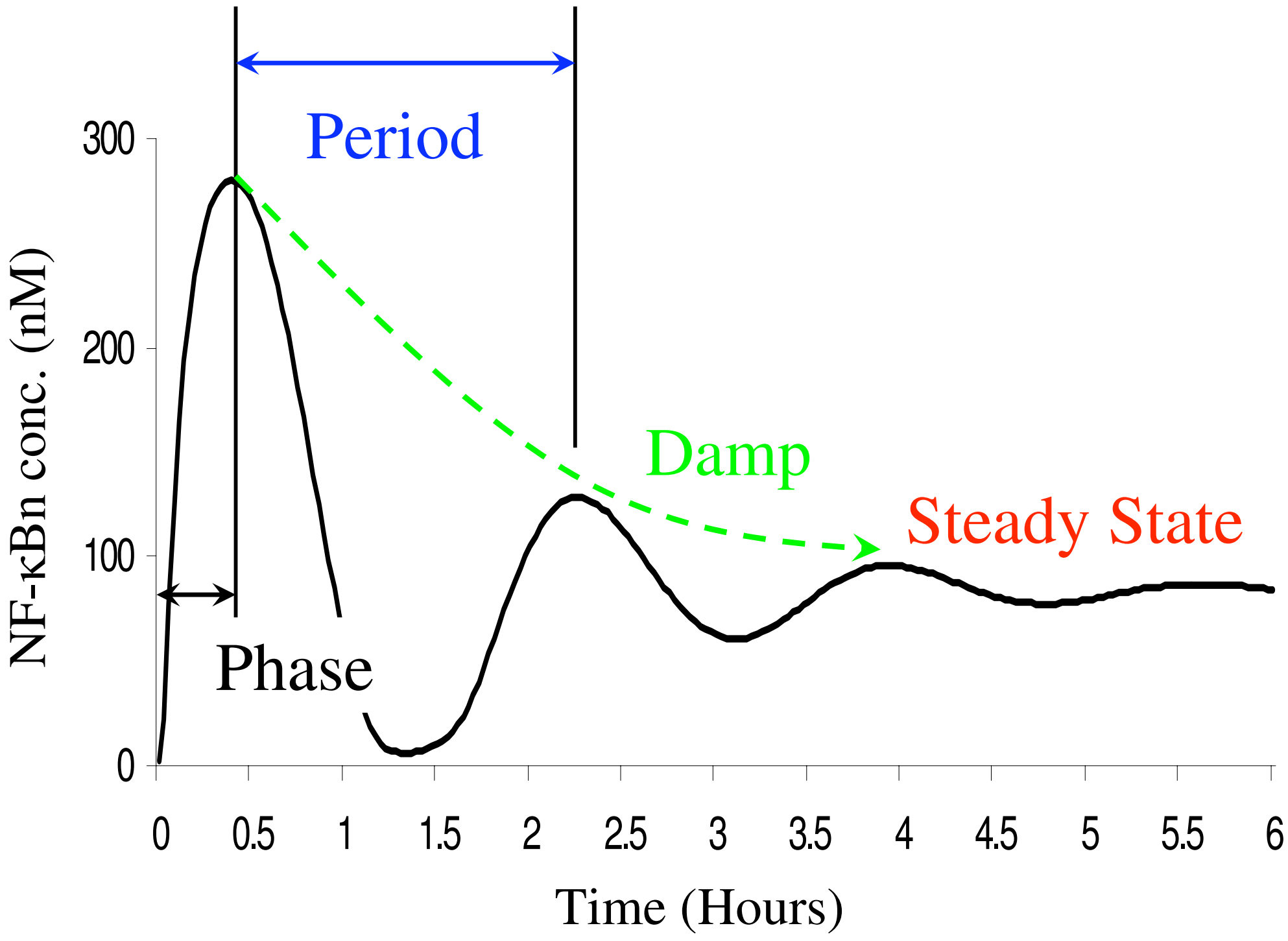

Figure 2

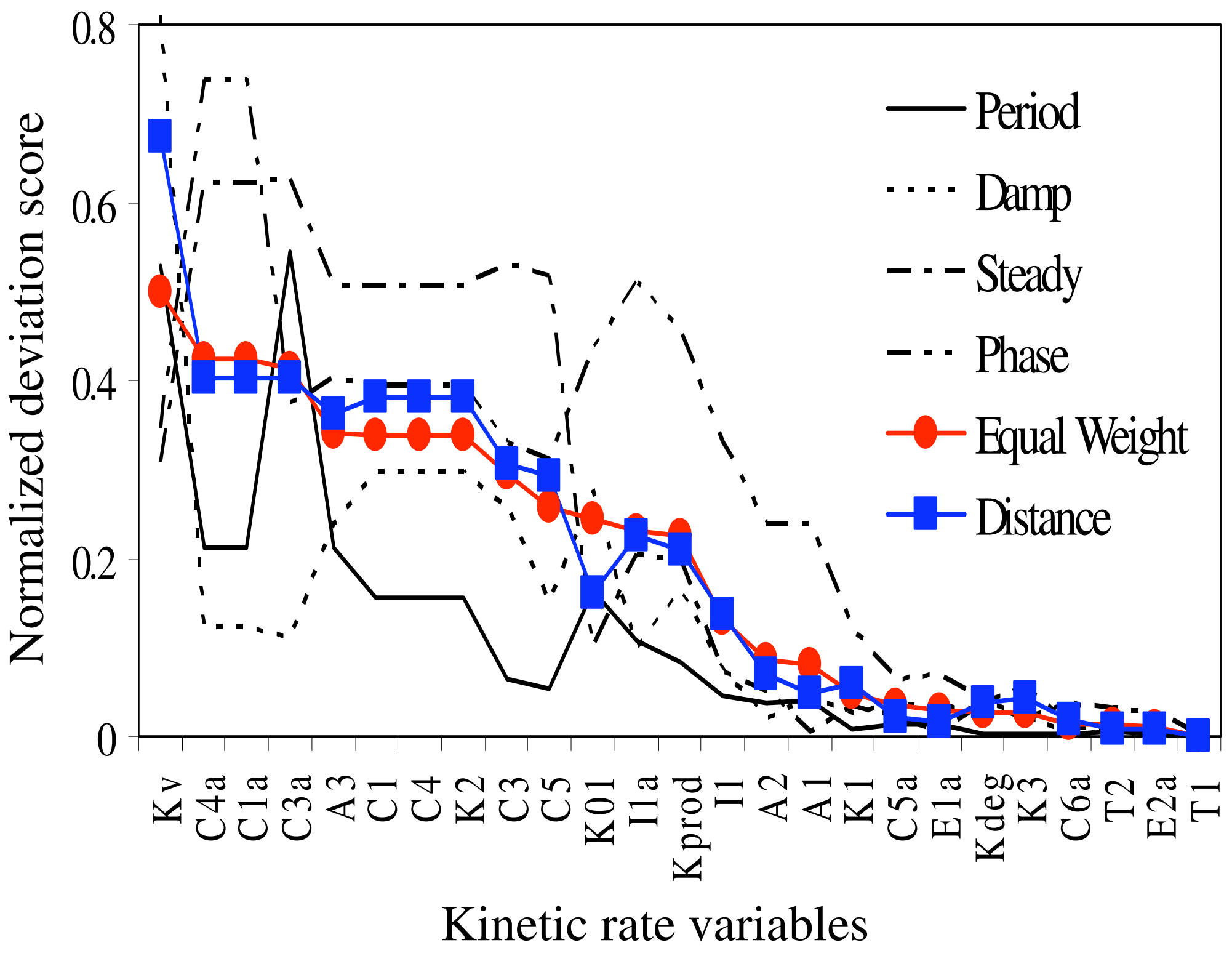

Figure 3

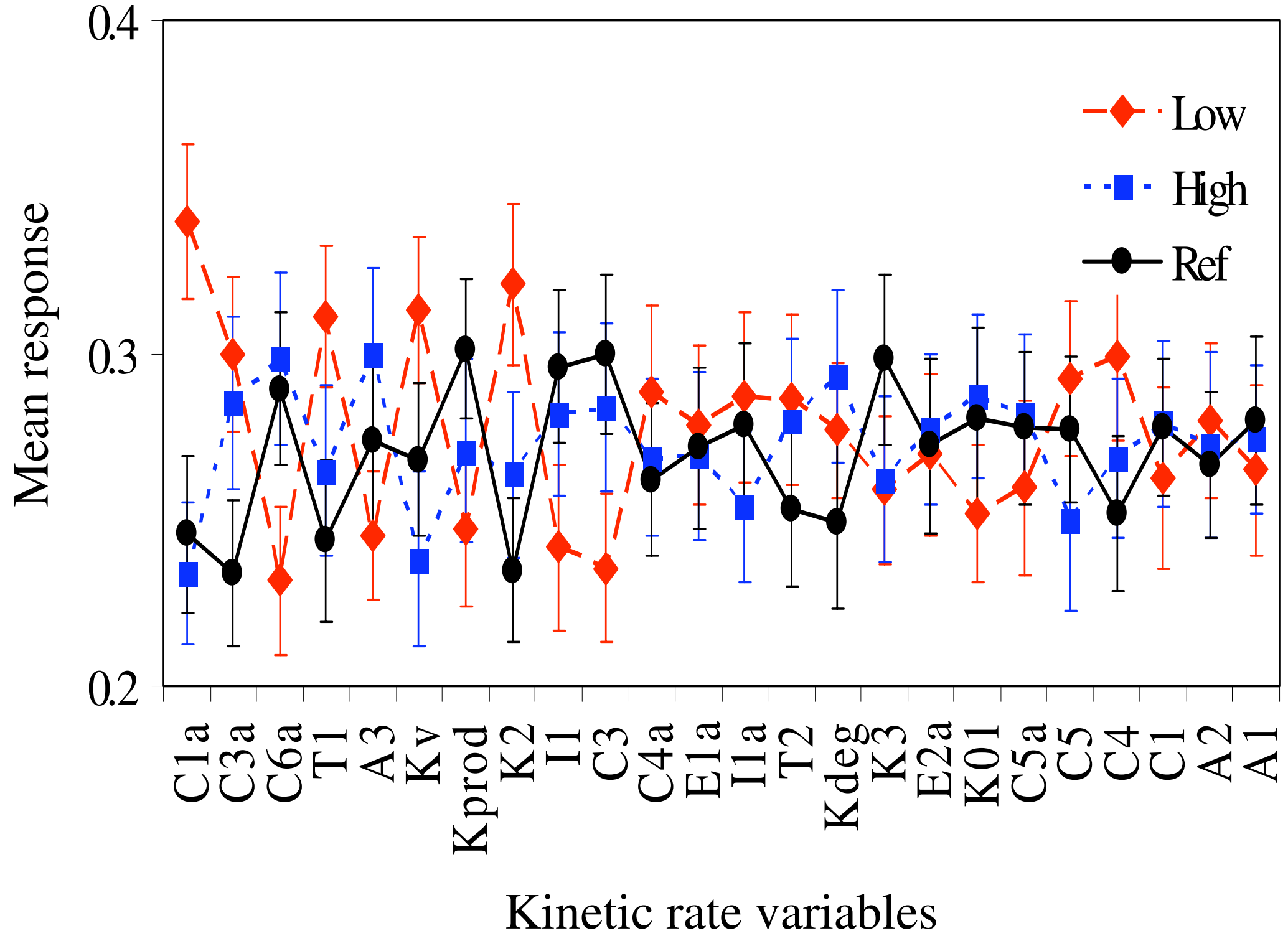

Figure 4

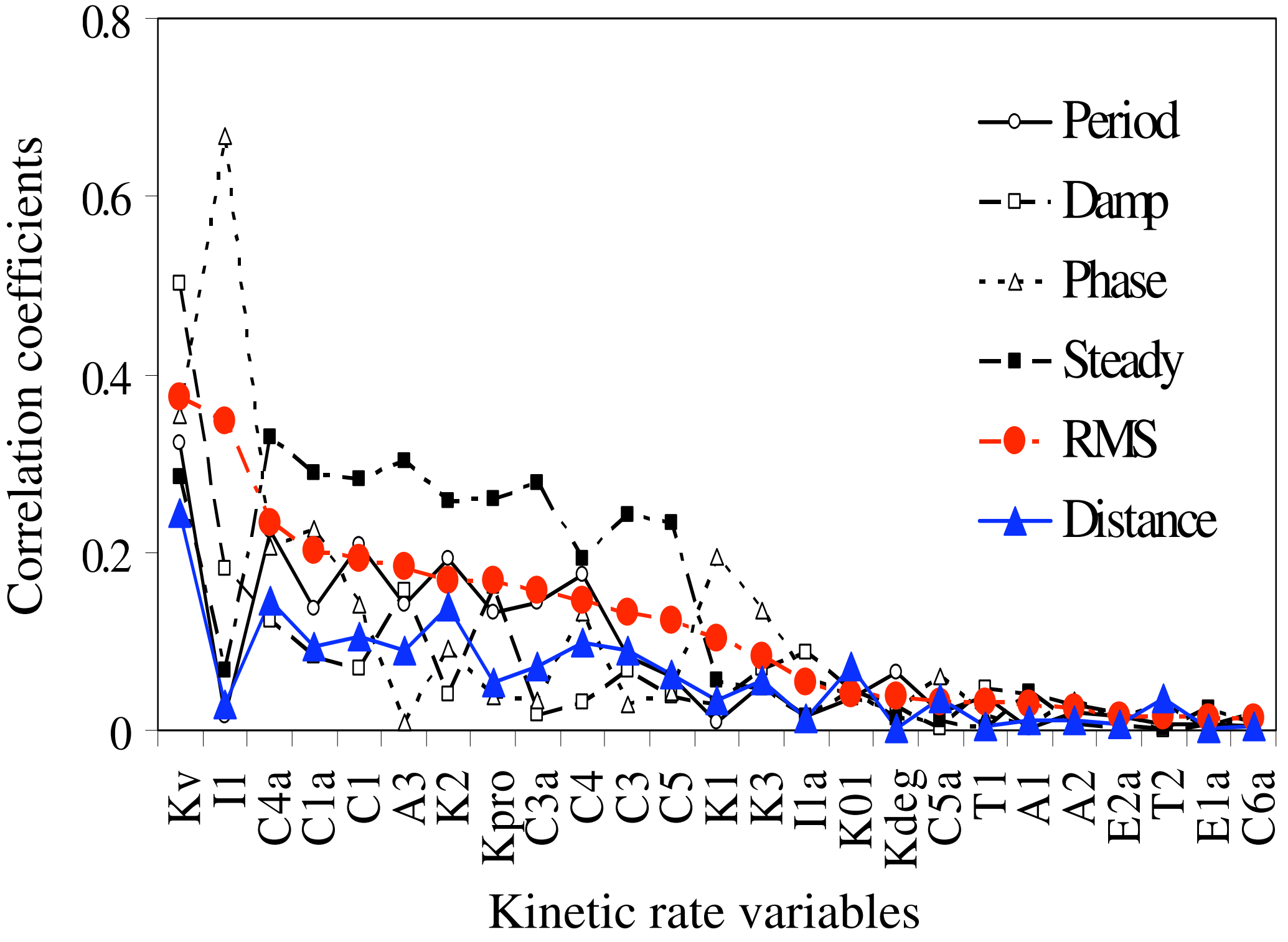

Figure 5

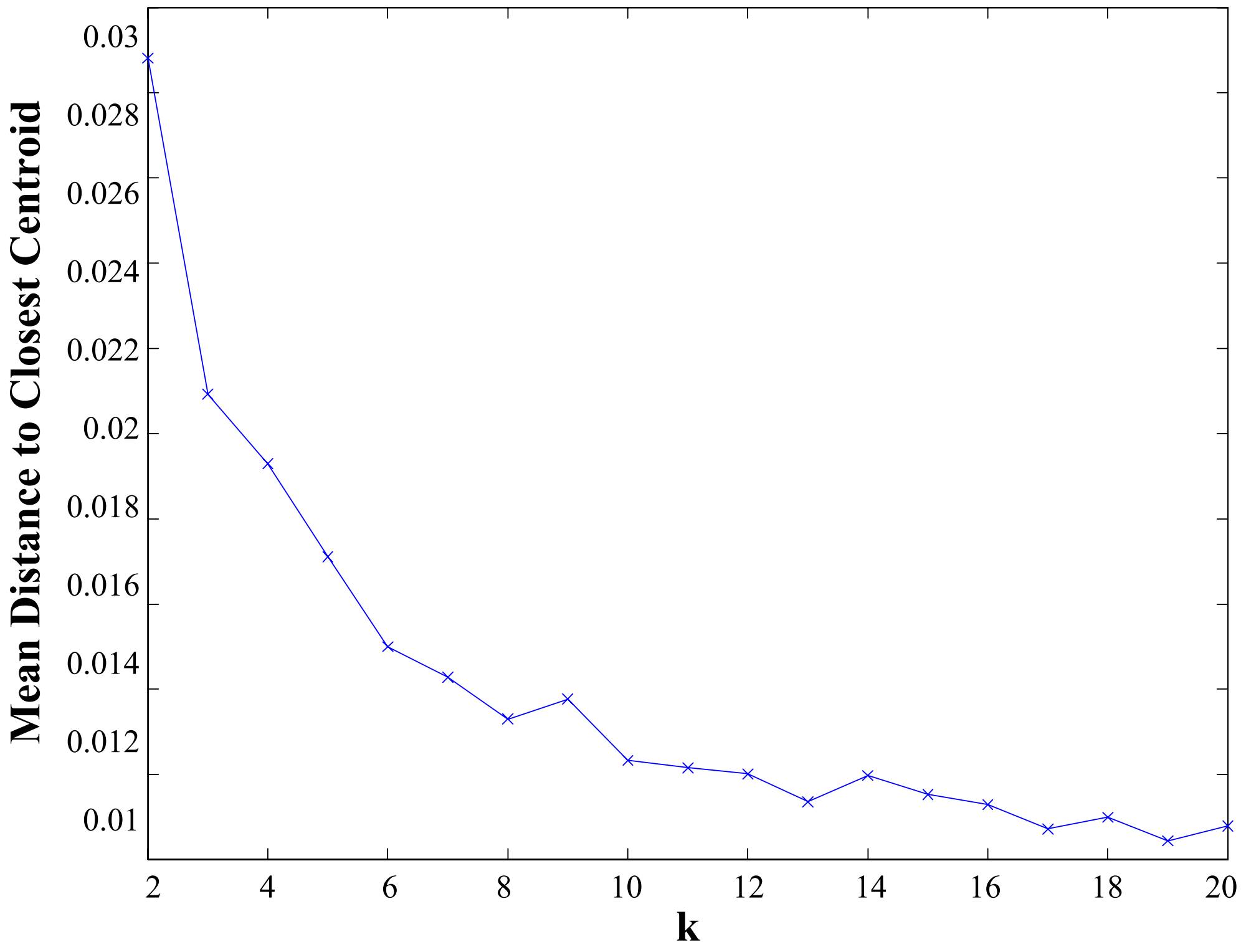

Figure 6

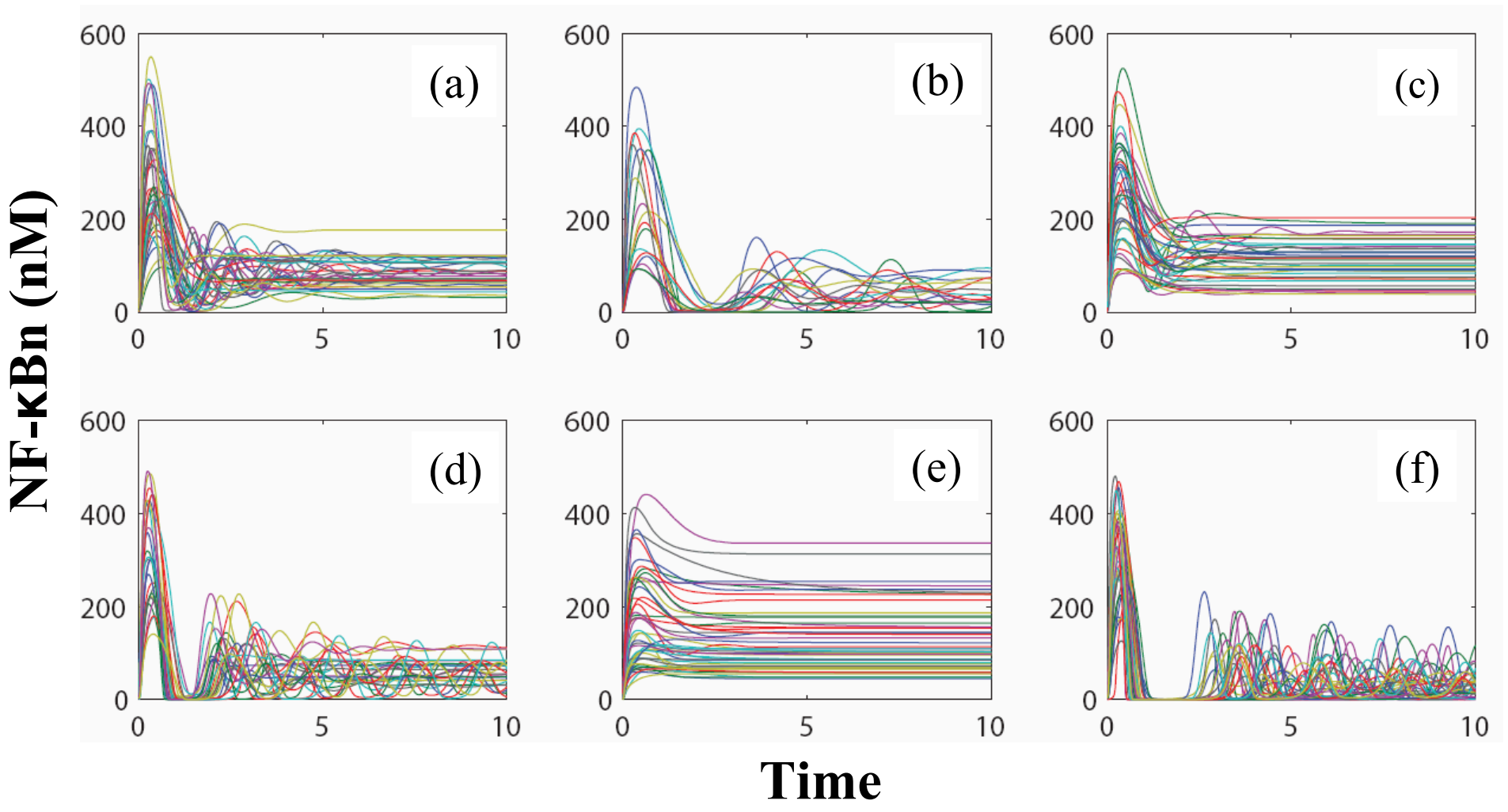

Figure 7

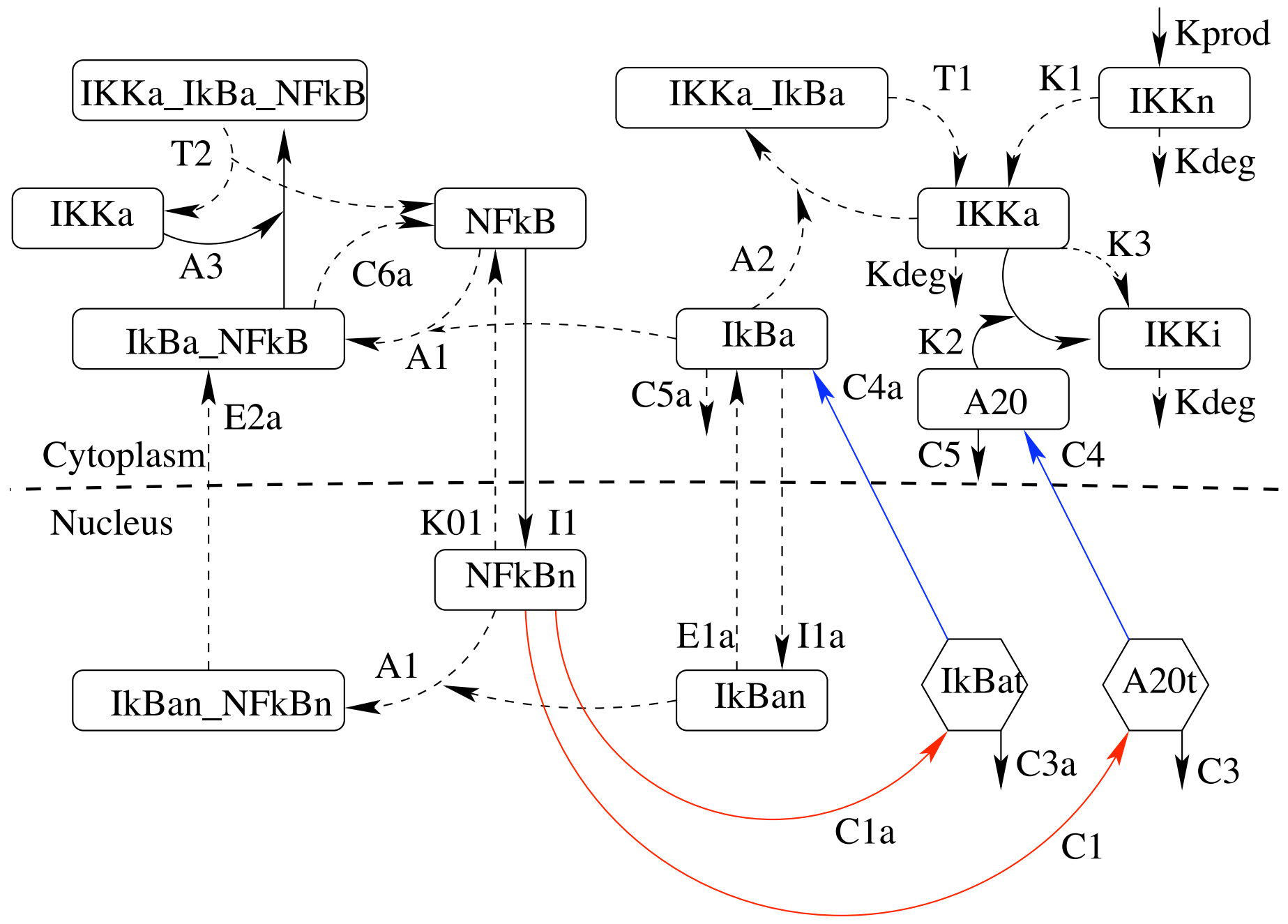

(a)

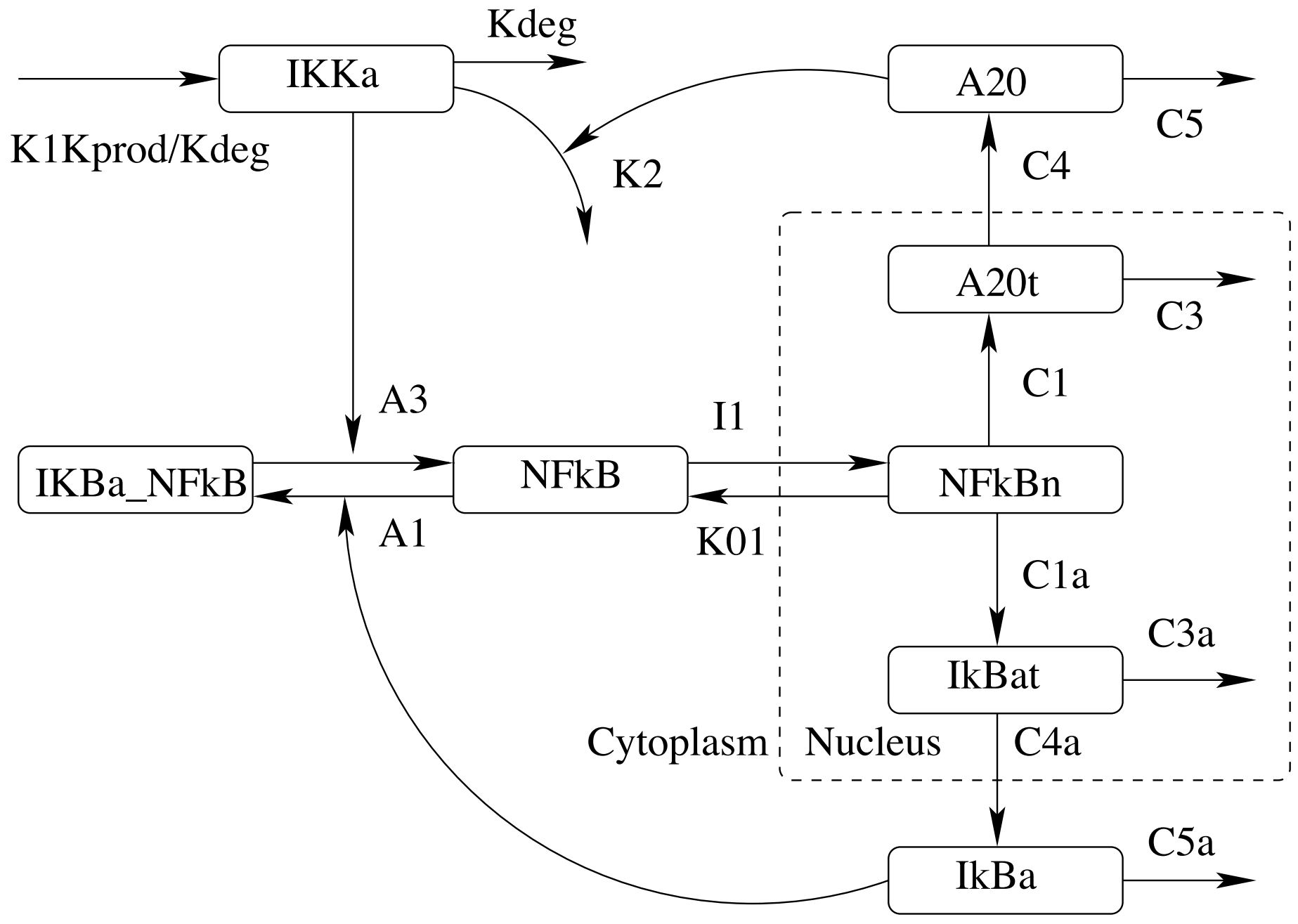

(b)

Figure 8